\documentclass[]{spie}  

\usepackage{amsmath,amsfonts,amssymb}
\usepackage{graphicx}
\usepackage[colorlinks=true, allcolors=blue]{hyperref}

\title{Improving accuracy of field work calculations for the analysis of the electron acceleration mechanisms in near-critical density plasma}

\author[a,b]{Diana Gorlova}
\author[a,b]{Ivan Tsymbalov}
\author[a]{Konstantin Ivanov}
\author[a]{Ekaterina Starodubtseva}
\author[a]{Andrei Savelev}

\affil[a]{Faculty of Physics, Lomonosov Moscow State University, Leninskye gory 1, 119991, Moscow,
Russia}
\affil[b]{Institute for Nuclear Research of the Russian Academy of Sciences, prospekt 60-letiya
Oktyabrya 7a, 117312, Moscow, Russia}

\authorinfo{Send correspondence to D.G.: E-mail: gorlova.da14@physics.msu.ru}

\begin{document} 
\maketitle

\begin{abstract}

It is demonstrated that the inherent structure of Boris pusher may result in the accumulation of errors in numerical integration when estimating the works of electric fields on particles. This, in turn, leads to an incorrect estimations of particle acceleration mechanisms. A method of error correction, reducing error to the acceptable levels of $<10\%$, is implemented. Numerical integration error in question is most pronounced in scenarios where substantial number of electrons are injected into the accelerating structure through wave breaking of plasma waves i.e. in near-critical density plasma. 

\end{abstract}


\section{INTRODUCTION}\label{sec:intro} 

Modern laser systems generate pulses of short femtosecond duration with peak powers up to several petawatts \cite{Gales2018-wu}. When focused, this allows one to achieve relativistic intensities ($10^{18}-10^{21}$ W/cm$^2$ for Ti:Sa laser systems) \cite{mourou2006optics}. The interaction of radiation of relativistic intensity with target atoms forms a laser plasma, which, can be used to efficiently accelerate electrons and ions \cite{Umstadter2003-tt}. The most commonly discussed electron acceleration mechanisms in relativistic laser plasma are laser wakefield acceleration (LWFA \cite{Gonsalves2019-ms}) and direct laser acceleration (DLA \cite{Pukhov1999-lf}), the first one resulting in electron beams with high energies (up to several GeV) and the second one - with high charge (up to $\mu$C \cite{Rosmej2020-lu}).

However, currently there is an interest in studying so-called "hybrid" acceleration regimes, where several electron acceleration mechanisms are present, for example LWFA to DLA regime \cite{Shaw2018-fz} or DLA to VLA (vacuum laser acceleration) \cite{Tsymbalov2019-qu}. There, in order to correctly establish acceleration mechanisms, an analysis of particle-in-cell (PIC) simulation results should be conducted; specifically one needs to calculate the work of various electric field components along the trajectories of test particles. Electron energy gain attributed to different components is explicitly related to various electron acceleration mechanisms (longitudinal field $E_x$ - LWFA acceleration, etc.). In this work, this procedure will be examined in more detail, and it will be shown that due to the inherent structure of the numerical scheme used in the PIC code an accumulation of integration errors occurs. A method of correcting this error will be proposed and implemented.

\section{NUMERICAL SIMULATION OF ELECTRON ACCELERATION}

\subsection{Simulation Setup}\label{sec:setup}

In the present work we used SMILEI PIC code to conduct the simulations \cite{DEROUILLAT2018351}. The calculations to establish the mechanisms of electron acceleration and the parameters of the electron beam were performed in 3D, while the simulations for electron beam deflection on the transverse gradient of electron concentration \cite{ivanov2023laser} were carried out in 2D3V mode due to the need for a large spatial domain. The problem under investigation is common for both simulation regimes, at it arises from type of electron's trajectory. Thus here, we will only highlight some common simulation parameters.

In all conducted calculations, a Gaussian laser pulse with a normalized vector potential of $a_0=1.5$ with a full width at half maximum (FWHM) duration of $\tau \textsubscript{FWHM}=50$ fs was focused to a spot with a diameter of $d\textsubscript{FWHM}=4$ $\mu$m. This corresponded to a vacuum intensity of approximately $I\approx 5\cdot 10^{18}$ W/cm², reflecting experimental conditions. At the beginning of the simulation, the target consisted of neutral carbon atoms, and their density profile was taken from the results of hydrodynamic modeling ($n_e \approx 0.05-0.2 n_{cr}$, detailed target parameters can be found in \cite{gorlova2023thz,ivanov2023laser}). In 3D simulations, 1 particle per cell was used due to their high computational complexity. For 2D simulations, 9 particles per cell were employed. Note, that 1 particle per cell for ions corresponds to 4-6 particles per cell for the electrons due to ionization. 

For subsequent analysis, during the simulations, the distributions of fields $E_x, E_y, E_z, B_z$, electron and ion concentrations $n_e$ and $n_i$, electron spatial and energy distributions, as well as trajectories of individual particles were recorded. Present work concerns problems of trajectory analysis specifically. All post-processing was carried out using Python scripts. In the discussion commonly used normalized quantities will be employed: spatial dimensions in units of the central wavelength of the laser pulse $\lambda$, time moments in units of $\tau=\lambda/c$, electron $n_e$ and ion $n_i$ concentrations in units of the critical electron concentration $n_{cr}$, field strengths $B$ and $E$ in units of the normalized vector potential $a_0 = \frac{eE}{\omega_L m c} = \frac{eB}{\omega_L m_e}$, where $e, m$ are the charge and mass of the electron, kinetic energies of electrons $E$ in units of the Lorentz factor $E = (\gamma-1)mc^2$, and field works $W$ in units of $mc^2 \approx 0.5$ MeV. Grid step sizes along the corresponding coordinates will be denoted as $\Delta x, \Delta y, \Delta z, \Delta\tau$.

\subsection{Calculation of the Work of Fields on an Electron Along its Trajectory}\label{sec:work_calc}

Full results on the acceleration process analysis will be published elsewhere, here we will discuss problem with work calculation that may arise during PIC simulation results analysis. As mentioned earlier, the most straightforward way to establish particle (here - electrons) acceleration mechanisms in laser-plasma acceleration setups is through the analysis of the works of various components of the electric field along the trajectories of electrons that acquired a significant momentum at the end of the calculation. Here only the electrons with final $p_x>5$ were analyzed. The work at each step $t=N\Delta \tau$ is calculated using the formula:

\begin{equation}
\label{eqn:general_work}
W_j=-e\int_0^t E_jv_j dt 
\end{equation}

where $j=x,y,z$ represents the projection onto the corresponding axis. During the analysis of numerical simulation results, the sum of works may deviate from the particle's energy, i.e.:

\begin{equation}
\label{eqn:work_bad}
W_x+W_y+W_z \neq \gamma +1
\end{equation}

ant the discrepancy can be significant ($\gg\gamma$). This inconsistency is primarily caused by the fact that in the commonly used Boris solver \cite{boris1970relativistic} (and many other solvers as well), the moments in time at which the electric fields $E_{x,y,z}$ (solving Maxwell's equations) and particle velocities $v_{x,y,z}$ (solving equations of motion) are determined are shifted by $\Delta\tau/2$. Therefore, the output data of the numerical simulation will also be determined at time points shifted by $\Delta\tau/2$, i.e., they will have a phase shift $\delta=\frac{\omega\Delta\tau}{2}$.  

\begin{figure}[ht]
\centering
\includegraphics[scale=0.58]{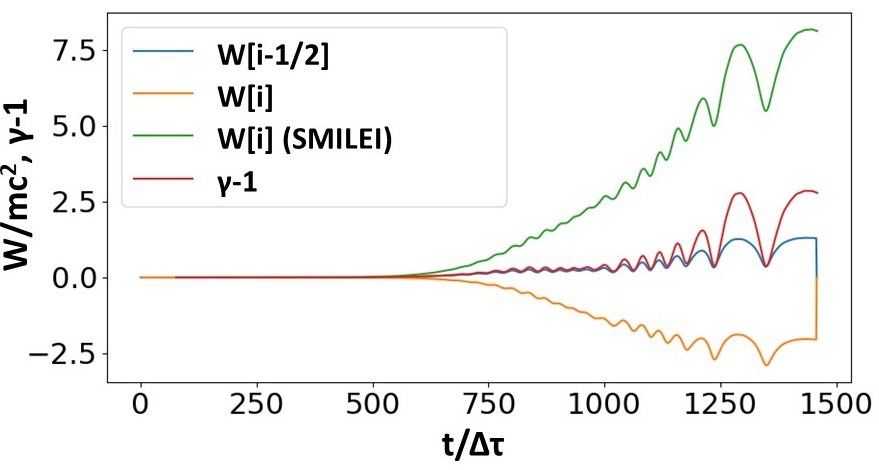}
\caption{The field works $W=W_x+W_y+W_z$, calculated directly from the values of $E$ and $v$ along the trajectory (yellow curve $W[i]$), with the field $E$ shifted during post-processing (blue curve $W[i-1/2]$), obtained from the beta version of the embedded SMILEI module (green curve $W[i]$ SMILEI), as well as $\gamma-1$ for a test particle as a function of time.
\label{img:work_discrepancy}}
\end{figure}

Let's illustrate why this phase shift will lead to errors in calculating the work done by fields using the formula \eqref{eqn:general_work}. Consider the motion of a single electron in the field of a plane wave as the first approximation of the interaction between the front of the laser pulse and electrons (i.e., when plasma waves have not yet formed). It can be shown that:

\begin{equation}
\label{eqn:work_no_discrepancy}
E \propto cos(\omega t) \quad v \propto sin(\omega t) \quad \rightarrow \quad Ev \propto sin(2\omega t) \quad \rightarrow \quad \int_0^\tau Evdt \propto cos(2\omega\tau)=0
\end{equation}
i.e., the average work done by the field over a period is zero. Now, let's introduce an additional phase shift $\delta$ between the field and the electron velocity, i.e.,

\begin{subequations}
\label{eqn:work_discrepancy}
\begin{gather}
E \propto \cos(\omega t) \quad v \propto \sin(\omega t+ \delta) \quad \rightarrow \quad Ev \propto \sin(2\omega t+\delta) + \sin(\delta) \\
\int_0^\tau Evdt \propto \frac{\cos(2\omega\tau+\delta)-\cos(\delta)}{2\omega}+\tau \sin(\delta)= \boxed{\tau \sin(\delta)\neq 0}
\end{gather}
\end{subequations}

i.e., the average work done by the field over a period is now not zero but depends on the period $\tau$ and the phase shift $\delta$. Obviously, when integrating over a time interval containing many periods, this constant component will accumulate linearly. Additionally, since the laser pulse is Gaussian, the field intensity $E$ rapidly increases at its front, further amplifying the error (Fig. \ref{img:work_discrepancy}, curve $W[i]$). Currently, an embedded method for calculating field works is being developed in the SMILEI code, which, at the moment, also leads to error accumulation (Fig. \ref{img:work_discrepancy}, curve $W[i]$ SMILEI).

Note that this issue is of concern only in the initial stages of particle acceleration. During the acceleration process and as the particle gains kinetic energy, the frequency of the electromagnetic field in the particle's co-moving frame will experience a Doppler shift into the red region. This means that both $\lambda$ and $\tau$ will significantly increase, causing the error accumulation to become negligible: $\delta \rightarrow 0, \tau \sin(\delta) \rightarrow const \ll\gamma$. Therefore, in scenarios where electron acceleration occurs to high energies ($\gamma\gtrapprox 20$, i.e. laser wakefield acceleration), the relative error may be small. 

\begin{figure}[ht]
\centering
\includegraphics[scale=0.5,page=1]{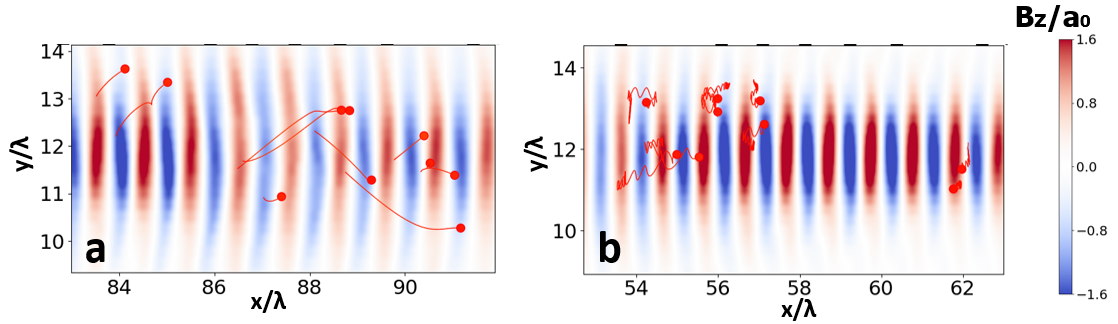}
\caption{Characteristic trajectories of electrons (red lines) and the magnetic field component of the laser pulse $B_z$ (red-blue) for electrons undergoing ionization injection (a) and injection due to the wave breaking of plasma waves (b).
\label{img:injection}}
\end{figure}

This issue is of big importance, however, for the acceleration regimes, concerned with generating electron beams with high charge and relatively low energies, i.e. most commonly - DLA in high density plasma ($n_e \gtrapprox 0.1 n_{cr}$), as in our case. In such interaction regimes two injection mechanisms are present: wave breaking of plasma waves (usually of backward stimulated Raman scattering (BSRS) origin) and ionization injection. In ionization injection, an electron, "breaking away" from an ion, immediately enters a region of strong field ($a_0>1$) and acquires a significant longitudinal velocity ($v_x/c > 0.5$, Fig. \ref{img:injection}a). In injection due to the wave breaking of plasma waves electron initially undergoes oscillations near the equilibrium position in the plasma wave ($v_x/c \approx 0$, Fig. \ref{img:injection}b) and acquires a significant longitudinal velocity only after the wave breaking. This prolonged electron' oscillation near the equilibrium position is what leads to fast error accumulation during trajectory analysis, i.e. error in work analysis is of importance for the interaction regimes where injection through the wave breaking is heavily present.

In the considered acceleration regime, the error accumulated at the pulse front routinely reaches $\frac{\mid W - (\gamma -1)\mid}{\gamma} \gg 1$, leading to an incorrect estimation of the field component works. The most straightforward and evident way to address this was the correction of the phase shift. This was done by linearly interpolating fields from time moments $t=n\Delta \tau$ to $t^{'}=(n-1/2)\Delta \tau$, where $n=0..N$ - time steps in the grid. For this, the fields $E_{x,y,z}$ were recorded at each time step $n$, and the values used for calculating works at this step were:

\begin{equation}
\label{eqn:e_shifred}
E_j[n-1/2] = \frac{E_j[n]+E_j[n-1]}{2}
\end{equation}

where $j=x,y,z$. The calculation of the work with the phase shift correction is shown in Fig. \ref{img:work_discrepancy} (curve $W[i-1/2]$). It can be observed that the discrepancy between $W$ and $(\gamma-1)$ is still present (likely due to interpolation inaccuracies), but the error stops accumulating over time.

\subsection{Comparison of Results for a Large Sample of Test Particles}\label{sec:comparison}

\begin{figure}[ht]
\centering
\includegraphics[scale=0.37,page=1]{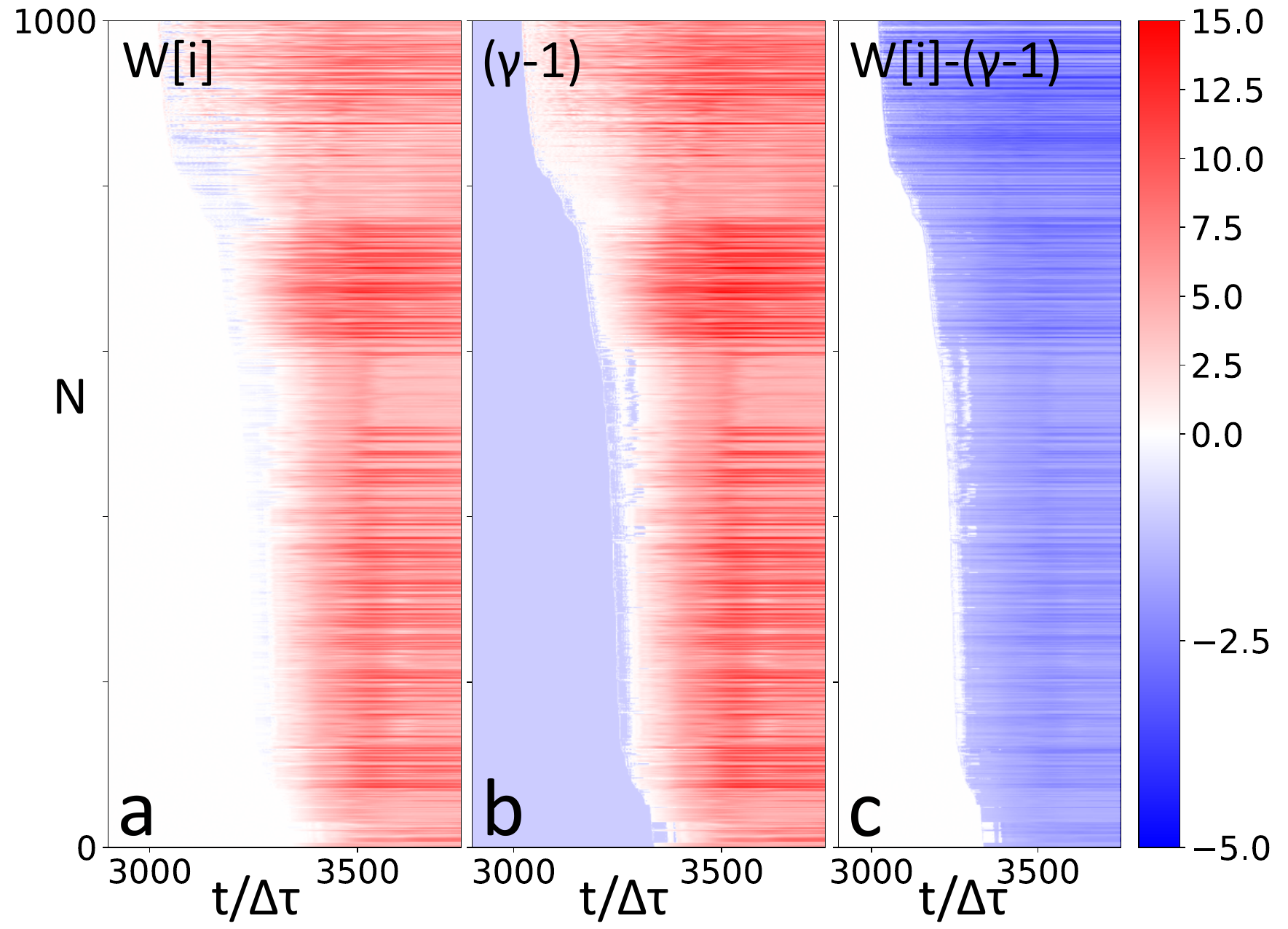}
\caption{The fields work $W=W_x+W_y+W_z$, calculated directly from the values of $E$ and $v$ (a, $W[i]$), $\gamma-1$ (b) and the divergence between them (c) for a sample of thousand of test particles as a function of time. All units are normalized in $mc^2$.
\label{img:work_before}}
\end{figure}

\begin{figure}[ht]
\centering
\includegraphics[scale=0.37,page=2]{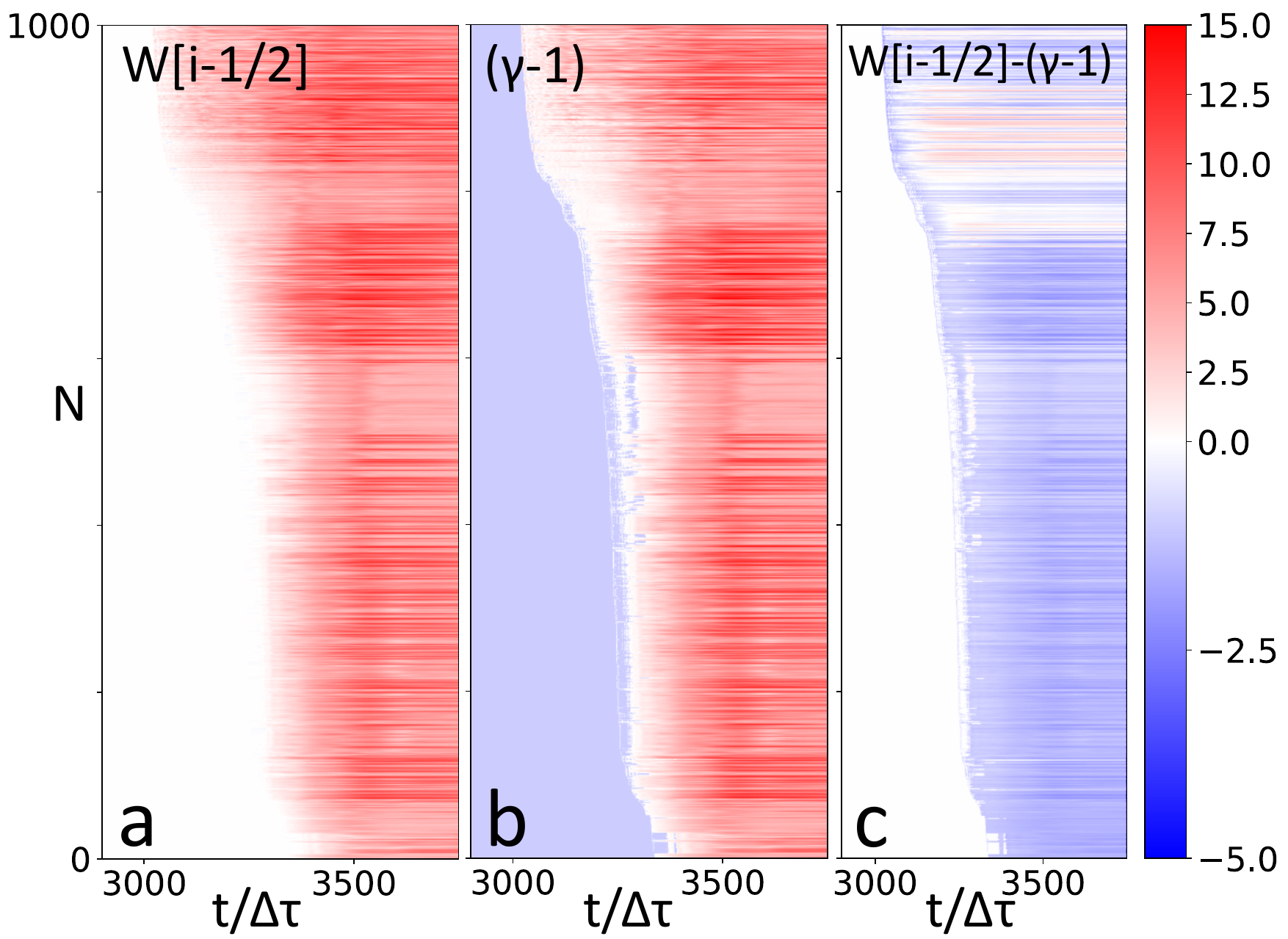}
\caption{The fields work $W=W_x+W_y+W_z$, calculated from the values of $E$ shifted during post-processing (a, $W[i-1/2]$), $\gamma-1$ (b) and the divergence between them (c) for a sample of thousand of test particles (same as Fig.\ref{img:work_before}) as a function of time. All units are normalized in $mc^2$.
\label{img:work_after}}
\end{figure}

With further acceleration of the test particles, the relative error $\frac{\mid W - (\gamma -1)\mid}{\gamma}$ decreases, and for most accelerated particles, it is less than $10\%$ by the end of the simulation. The results of a works calculation for a large sample of test particles are presented in Fig. \ref{img:work_before} and Fig. \ref{img:work_after}. It can be observed that in the absence of correction in the field work calculation (Fig. \ref{img:work_before}), the divergence between the particle kinetic energy and the sum of the field works at the end of the simulation is significant and can reach up to $W[i]-(\gamma-1) \approx -5$ (this value can be both negative and positive depending on the initial $\delta$, i.e. determined by solver and code properties). There is also some non-physical behavior present: at the beginning stages of acceleration $W[i]<0$, while $\gamma-1\geq 0$ condition is always met. The use of the algorithm of error correction described above leads to a significant reduction in this divergence (see Fig. \ref{img:work_after}). For most test particles at the end of the simulation the condition $\mid W[i-1/2]-(\gamma-1)\mid<1 $ is met, and non-physical behavior does not arise. Further improvements in the fields interpolation scheme may reduce the discrepancy (\ref{eqn:work_bad}) to the required level, we found the $10\%$  level satisfactory for further analysis.

\section{CONCLUSIONS}

It has been demonstrated that the inherent structure of Boris pusher, commonly used in PIC codes, may result in the accumulation of errors in numerical integration when estimating the works of electric fields on particles. This, in turn, leads to an incorrect estimations of particle acceleration mechanisms. A method of error correction based on fields interpolation during post-processing has been proposed and successfully implemented, reducing $\mid\frac{W}{\gamma-1}\mid$ to the acceptable levels of $<10\%$. This divergence effect is most pronounced in scenarios where the electron trajectory involves a significant amount of oscillations near the equilibrium position, i.e. when substantial number of electrons are injected into the accelerating structure through wave breaking of plasma waves (usually in near-critical density plasma). Nevertheless, the results demonstrated in this work are general as they are caused by the structure of the Boris numerical scheme itself. 

\acknowledgments 
 
This work was supported by RSCF Grant №22-79-10087.  D.G. and E.S. acknowledge the Foundation for Theoretical Research ‘BASIS’ for financial support.

\bibliography{report} 

\begin{thebibliography}{10}

\bibitem{Gales2018-wu}
Gales, S., Tanaka, K., Balabanski, D., Negoita, F., Stutman, D., Tesileanu, O., Ur, C., Ursescu, D., Andrei, I., Ataman, S., et~al., ``The extreme light infrastructure—nuclear physics (eli-np) facility: new horizons in physics with 10 pw ultra-intense lasers and 20 mev brilliant gamma beams,'' {\em Reports on Progress in Physics}~{\bf 81}(9),  094301 (2018).

\bibitem{mourou2006optics}
Mourou, G.~A., Tajima, T., and Bulanov, S.~V., ``Optics in the relativistic regime,'' {\em Reviews of modern physics}~{\bf 78}(2),  309 (2006).

\bibitem{Umstadter2003-tt}
Umstadter, D., ``Relativistic laser-plasma interactions,'' {\em J. Phys. D Appl. Phys.}~{\bf 36}(8),  R151--R165 (2003).

\bibitem{Gonsalves2019-ms}
Gonsalves, A.~J., Nakamura, K., Daniels, J., Benedetti, C., Pieronek, C., De~Raadt, T. C.~H., Steinke, S., Bin, J.~H., Bulanov, S.~S., Van~Tilborg, J., Geddes, C. G.~R., Schroeder, C.~B., T{\'o}th, C., Esarey, E., Swanson, K., Fan-Chiang, L., Bagdasarov, G., Bobrova, N., Gasilov, V., Korn, G., Sasorov, P., and Leemans, W.~P., ``Petawatt laser guiding and electron beam acceleration to 8 {GeV} in a {Laser-Heated} capillary discharge waveguide,'' {\em Phys. Rev. Lett.}~{\bf 122}(8),  84801 (2019).

\bibitem{Pukhov1999-lf}
Pukhov, A., Sheng, Z.~M., and Meyer-ter Vehn, J., ``Particle acceleration in relativistic laser channels,'' {\em Phys. Plasmas}~{\bf 6}(7),  2847--2854 (1999).

\bibitem{Rosmej2020-lu}
Rosmej, O.~N., Gyrdymov, M., G{\"u}nther, M.~M., Andreev, N.~E., Tavana, P., Neumayer, P., Z{\"a}hter, S., Zahn, N., Popov, V.~S., Borisenko, N.~G., Kantsyrev, A., Skobliakov, A., Panyushkin, V., Bogdanov, A., Consoli, F., Shen, X.~F., and Pukhov, A., ``High-current laser-driven beams of relativistic electrons for high energy density research,'' {\em Plasma Phys. Controlled Fusion}~{\bf 62}(11),  115024 (2020).

\bibitem{Shaw2018-fz}
Shaw, J.~L., Lemos, N., Marsh, K.~A., Froula, D.~H., and Joshi, C., ``Experimental signatures of direct-laser-acceleration-assisted laser wakefield acceleration,'' {\em Plasma Phys. Controlled Fusion}~{\bf 60}(4),  44012 (2018).

\bibitem{Tsymbalov2019-qu}
Tsymbalov, I., Gorlova, D., Shulyapov, S., Prokudin, V., Zavorotny, A., Ivanov, K., Volkov, R., Bychenkov, V., Nedorezov, V., Paskhalov, A., Eremin, N., and Savel'ev, A., ``Well collimated {MeV} electron beam generation in the plasma channel from relativistic laser-solid interaction,'' {\em Plasma Phys. Controlled Fusion}~{\bf 61},  075016 (July 2019).

\bibitem{DEROUILLAT2018351}
Derouillat, J., Beck, A., Pérez, F., Vinci, T., Chiaramello, M., Grassi, A., Flé, M., Bouchard, G., Plotnikov, I., Aunai, N., Dargent, J., Riconda, C., and Grech, M., ``Smilei : A collaborative, open-source, multi-purpose particle-in-cell code for plasma simulation,'' {\em Computer Physics Communications}~{\bf 222},  351--373 (2018).

\bibitem{ivanov2023laser}
Ivanov, K., Gorlova, D., Tsymbalov, I., Tsygvintsev, I., Shulyapov, S., Volkov, R., and Savelev, A., ``Laser-driven pointed acceleration of electrons with preformed plasma lens,'' {\em arXiv preprint arXiv:2309.10530}  (2023).

\bibitem{gorlova2023thz}
Gorlova, D., Tsymbalov, I., Tsygvintsev, I., and Savelev, A., ``Thz transition radiation of electron bunch laser-accelerated in long-scale near-critical density plasmas,'' {\em arXiv preprint arXiv:2310.19282}  (2023).

\bibitem{boris1970relativistic}
Boris, J.~P. et~al., ``Relativistic plasma simulation-optimization of a hybrid code,'' in [{\em Proc. Fourth Conf. Num. Sim. Plasmas}{\nolinebreak\hspace{0.1em}]},   3--67 (1970).

\end{thebibliography}
\bibliographystyle{spiebib} 

\end{document}